\documentstyle[psfig]{lamuphys}
\makeatletter
\let\chapter\hid@chapter
\makeatother
\begin{document}
\pagenumbering{arabic}
\title{The intrinsic properties of the local interstellar medium}

\author{Olivia Puyoo\inst{1} and Lotfi Ben\,Jaffel\inst{1}}

\institute{Institut d'Astrophysique de Paris, 98 bis Blvd Arago, 75014 Paris,
France}

\maketitle

\begin{abstract}
We propose a new method to constrain the actual state of the interstellar
 cloud that surrounds the solar system. 
Using Voyager UVS Lyman-$\alpha$ sky maps and the powerful principle of
 invariance, we derive the H distribution all 
along the spacecraft path. Provided current models of 
the heliopause interface between the solar and the interstellar winds, 
we extrapolate this 
distribution to farther distances from the Sun and infer in a 
self consistent way key parameters of the local cloud. 
Our findings are a high 
interstellar hydrogen density of $\sim0.24\,{\rm cm^{-3}}$ 
and a weak 
ionization $\frac{{\rm n(H^+)}}{{\rm n(H^+)+n(H)}}\simeq\,14\%$.
\end{abstract}
\section{Introduction}
It is widely believed that the local interstellar medium 
represents a point sample 
of the more extended interstellar medium (ISM), and that 
its studies through its
interplanetary signature is of particular interest to understand the origin
and the evolution of the ISM. 
Because of the relative motion of the Sun with respect to the local 
interstellar cloud (LIC), 
the neutral component of the interstellar wind
penetrates deeply inside the solar system, where it becomes accessible to
in situ detections. After its discovery in the early 70's 
in the Earth neighborhood, interpretation of backscattered H Lyman-$\alpha$ and
HeI $58.4\ {\rm nm}$ solar radiations has been extensively employed to derive
H and He abundances in the local cloud. 
As classical interpretations of UV sky maps usually require the modeling
of the H distribution in the heliosphere with the problem of filtration at the
heliopause and the calculation of the radiation field including 
multi-scattering effects, the derivation of neutral abundances proved to be
difficult, and the values found were consequently poorly constrained.\\
The aim of this paper is to present 
a new technique for deriving the intrinsic properties of the
LIC by interpreting differently the UV backscattered radiation
measurements. This method consists in constraining the H distribution stage
by stage, starting from the Earth neighborhood up to the far unperturbed
LIC. In a first step, we determine the H distribution along the 
Voyager trajectory by applying the principle of invariance to UVS
Lyman-$\alpha$ sky maps recorded for several positions of the spacecraft 
(Puyoo et al 1997). 
Then we connect this distribution to the LIC one
through a model of interaction between the solar and the interstellar
flows. Lastly, assuming the LIC in a local steady state, we derive 
the intrinsic properties of the local cloud, i.e. the
densities of both neutral and ionized Hydrogen and Helium 
in a self consistent way.
\section{Method}
Our technique is a continuation of a method recently proposed by Puyoo
et al (1997) to derive in a self-consistent way the H neutral density in the
inner heliosphere. 
This method is based on the invariance principle, which expresses, for a
slab, the direct relationship between the radiation field incident on its
boundaries and its optical properties, i.e. the reflexion and 
transmission coefficients and the contribution of inner sources. 
As the coefficients are dependent on the scattering gas distribution, the H
distribution within the layer is therefore directly accessed. 
With that in view, we apply our technique 
to VOYAGER/UVS \hbox{Lyman-$\alpha$} sky maps taken at different
locations of the Spacecraft from 4 to 35 AU, 
which provides the H density distribution 
as shown in Fig. 1. 
It is worth to note
that by using the principle of invariance, 
we avoided the modeling of the whole complex medium as it was
previously required by classical techniques 
to interpret Voyager data. However, 
due to the local nature of our method, a limitation immediately appears as 
we can only constrain
the density for regions visited by the Voyager spacecraft and for which
Ly-$\alpha$ sky maps exist.\\
\begin{figure}[thbp]
\psfig{file=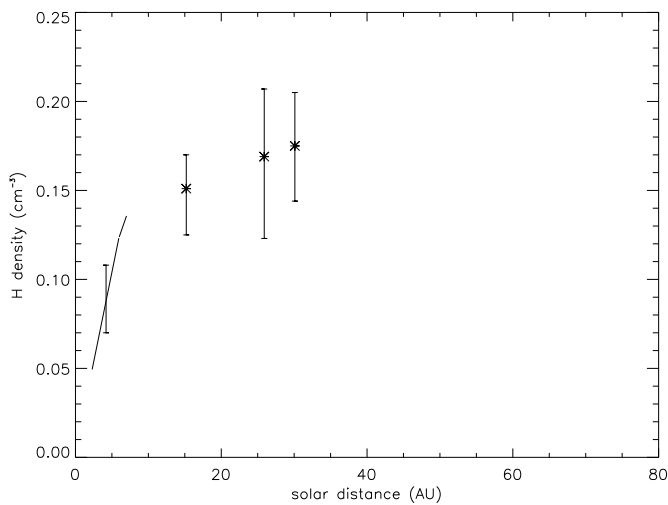,width=10cm,clip=}
\caption[]{H density distribution along Voyager 2 path derived from
Lyman-$\alpha$ sky maps using the invariance principle.}
\label{p1}
\end{figure}
To reach larger distances from the Sun, 
we extrapolate the distribution shown in Fig. 1, using 
the sophisticated model of interaction between the solar wind and the
supersonic flow as developed by Baranov \& Malama (1993). In such models, the
filtration effects due to charge exchanges between interstellar hydrogen
atoms and
protons at the heliopause depend mainly on two parameters: the Mach
number and the proton density of the LIC. In the following, 
we fix the Mach number at its
commonly assumed value M$\simeq\,2.$ (Baranov \& Malama 1993)
and consider
the LIC proton density as a free parameter. We then extrapolate Voyager  H
distribution shown in Fig. 1 by varying both the LIC
proton (${\rm n_{H^+}}$) and the H neutral (${\rm n_{H\infty}}$) 
densities, which provides 
a first relation between the
two parameters (see Fig. 2). A second relation is however necessary to 
simultaneously 
constrain them. For that end, 
we assume the unperturbed LIC to be in a
local statistical equilibrium, and we solve
the charge balance equations (e.g. Vallerga \& Welsh 1993) controlled
by the ionization due to cosmic rays, known EUV stellar sources and the
estimated contribution of the conductive interface between the local cloud and
the surrounding local bubble (Slavin 1989).
\begin{figure}[thbp]
\psfig{file=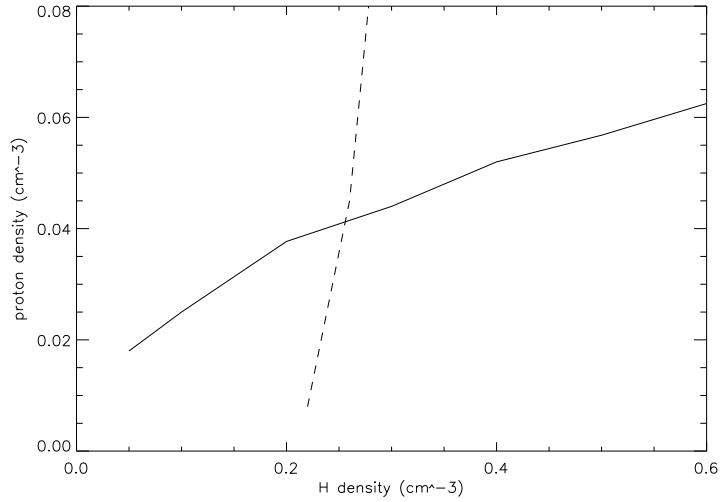,width=10cm,clip=}
\caption[]{Relations between the LIC proton and H neutral densities derived 
from the connection of the LIC to the Voyager H distribution (see Fig. 1)
through 
Baranov model (dashed curve) and from the assumption of a local ionization 
equilibrium, taking $\Gamma_{\rm H}\simeq\,2.5\,10^{-15}\,s^{-1}$ and
$\Gamma_{\rm He}\simeq\,6.9\,10^{-15}\,s^{-1}$ (solid line). 
The intersection gives the actual H and proton 
densities, 
${\rm n_{H\infty}}\simeq\,0.24\,{\rm cm^{-3}}$ 
${\rm n_{H^+}}\simeq\,0.043\,{\rm cm^{-3}}$ respectively, in the local cloud.}
\label{}
\end{figure}
With the additional assumption of a cosmic ratio of H/He=10, we obtain a
second relation between proton and H densities.\\
As shown in
Fig. 2, this relation, 
combined with the one deduced through the Voyager
distribution and the Baranov model, allows to derive 
in a self consistent way the following set of properties of the local cloud
at $\sim$1000 AU from the Sun:
\begin{itemize}
\item H neutral: ${\rm n_H}=0.24\pm\,0.05\,{\rm cm^{-3}}$
\item proton: ${\rm n_{H^+}}=0.043\pm\,0.005\,{\rm cm^{-3}}$
\item He neutral: ${\rm n_{He}}=(2.7\pm\,0.5)10^{-2}\,{\rm cm^{-3}}$
\item He$^+$: ${\rm n_{He^+}}=(1.5\pm\,0.3)\,10^{-3} \,{\rm cm^{-3}}$
\item electron: ${\rm n_e}=0.044\pm\,0.005\,{\rm cm^{-3}}$
\item He$^{++}$: ${\rm n_{He^{2+}}}=(1.0\pm .2)\,10^{-5}\,{\rm cm^{-3}}$
\end{itemize}                       
\section{conclusion}
After extraction of the hydrogen distribution from the Voyager 
Ly-$\alpha$ sky maps 
inside the heliosphere (see Fig. 1) and assuming local statistical
equilibrium, we derived that the 
LIC has a high H density ${\rm n_{H\infty}}\simeq\,0.24\,{\rm cm^{-3}}$ 
and is weakly
ionized with an ionization fraction near the Sun of $\sim\,15\%$ and
$\sim\,6\%$ for H and He respectively. 
The self-consistency of the method used here, makes the derived set of
LIC parameters unique for the ionization sources considered here, which
could however 
not explain the enhanced He ionization with respect to H found by Dupuis 
et al (1995).\\
In order to check the validity of the weak ionization inferred for H, it is
interesting to compare our electron density to available measurements. We 
apply a simple model that calculates 
the photoionization of H and He from the Sun out to the edge of the local 
cloud to 
different lines of sight. Toward $\epsilon$Cma or Capella, we 
obtain a mean electron density $<{\rm n_e}>\simeq\,0.08\,{\rm cm^{-3}}$, which 
is in good agreement with recent HST measurements made by respectively 
Gry et al (1995) 
and Wood \& Linsky (1995). As far as the ionization of H is the main
pool for electron production, this result seems to confirm that this
process is well described by our model. Nevertheless, an additional source
that preferentially ionizes He atoms is required to explain EUVE results
(Vallerga 1996).

%
%

\end{document}